\def\cm{{\rm\thinspace cm}}

\def\erg{{\rm\thinspace erg}}

\def\keV{{\rm\thinspace keV}}
\def\km{{\rm\thinspace km}}

\def\Msun{\hbox{$\rm\thinspace M_{\odot}$}}

\def\s{{\rm\thinspace s}}
\def\rg{{\rm\thinspace r$_{\rm g}$}}

\def\sr{{\rm\thinspace sr}}

\def\ergpcmsqps{\hbox{$\erg\cm^{-2}\s^{-1}\,$}}

\def\ergps{\hbox{$\erg\s^{-1}\,$}}

\def\kmps{\hbox{$\km\s^{-1}\,$}}

\def\spose#1{\hbox to 0pt{#1\hss}}
\def\approxlt{\mathrel{\spose{\lower 3pt\hbox{$\sim$}}
        \raise 2.0pt\hbox{$<$}}}
\def\approxgt{\mathrel{\spose{\lower 3pt\hbox{$\sim$}}
        \raise 2.0pt\hbox{$>$}}}

\documentclass{article}
\usepackage{psfig}
\usepackage{aaspp4}

\usepackage{psfig}

\begin{document}
\pagestyle{empty}

\title{Broad iron lines in Active Galactic Nuclei}
\author{ A.C. Fabian\altaffilmark{1}, K. Iwasawa\altaffilmark{1}, 
C.S. Reynolds\altaffilmark{2,3}, A.J. Young\altaffilmark{4}}

\altaffiltext{1}{Institute of Astronomy, Madingley Road, Cambridge CB3 0HA
UK}
\altaffiltext{2}{JILA, University of Colorado, Boulder CO~80309}
\altaffiltext{3}{Hubble Fellow}
\altaffiltext{4}{Department of Astronomy, University of Maryland,
College Park MD~20742}

\begin{abstract}
An intrinsically narrow line emitted by an accretion disk around a black
hole appears broadened and skewed as a result of the Doppler effect and
gravitational redshift. The fluorescent iron line in the X-ray band at
6.4 -- 6.9~keV is the strongest such line and is seen in the X-ray
spectrum of many active galactic nuclei and, in particular, Seyfert
galaxies. It is an important diagnostic with which to study the geometry
and other properties of the accretion flow very close to the central
black hole. The broad iron line indicates the presence of a standard
thin accretion disk in those objects, often seen at low inclination. The
broad iron line has opened up strong gravitational effects around black
holes to observational study with wide-reaching consequences for both
astrophysics and physics.
\end{abstract}

\section{Introduction}

In recent years, observations with both ground-based and space-based
instruments have led to realization that most, if not all, nucleated
galaxies harbor a massive black hole at their center (Kormendy \&
Richstone 1995; Magorrian et al. 1998).  While many of these black
holes appear to be relatively isolated, some fraction accrete
significant amounts of material from the surrounding galaxy.  The
angular momentum of the incoming material leads to the formation of a
flattened rotating disk --- the accretion disk.  The gravitational
potential energy of material flowing through the accretion disk is
converted into radiative (i.e. electromagnetic) and kinetic energy.
These powerful and compact energy sources, observed in approximately
1--10\% of galaxies are termed {\it active galactic nuclei} (AGN).
AGN are also observed to be copious X-ray emitters.  These X-rays are
thought to originate from the innermost regions of an accretion disk
around a central supermassive black hole.  Since the accretion disk
itself is expected to be an optical/UV emitter, the most likely
mechanism producing the X-rays is inverse Compton scattering of these
soft photons in a hot and tenuous corona that sandwiches the accretion
disk.  Thus, in principle, the study of these X-rays should allow the
immediate environment of the accreting black hole as well as the
exotic physics, including strong-field general relativity, that
operates in this environment to be probed.

This review discusses how, in the past decade, X-ray astronomy has begun
to fulfill that promise.  Guided by observations with the {\it Ginga},
ASCA, RXTE and BeppoSAX satellites, there is a broad concensus that
X-ray irradiation of the surface layers of the accretion disk in a class
of AGN known as Seyfert 1 galaxies gives rise to fluorescent K$\alpha$
emission line of cold iron via the process of ``X-ray reflection''.
Since this line is intrinsically narrow in frequency, the observed
energy profile of the line is shaped by both special relativistic
(i.e. Doppler shifting) and general relativistic (i.e.  gravitational
redshifting and light bending) effects into a characteristic skewed
profile predicted over a decade ago (Fabian et al 1989) and first
clearly seen in ASCA data (Tanaka et al 1995).  Since these lines are
typically broadened to a full-width half maximum of $5\times 10^4\kmps$
or more, they are often referred to as ``broad iron lines''. After
discussing the physical processes responsible for the production of
these spectral signatures, we will summarize the current observational
status of broad iron line studies.  We will show how current
observations are already addressing the nature of the accretion disk
within a few gravitational radii of the black hole.  Observations of the
broad iron line also provide valuable insights into the physical
differences behind AGN of differing luminosities and type.  Finally, we
discuss and attempt to predict the results that will emerge from high
throughput X-ray spectroscopy with {\it XMM--Newton}, {\it
Constellation-X} and {\it XEUS}. We argue that these future data will
provide unprecedented constraints on the spacetime geometry near the
black hole (thereby measuring the spin of the black hole), well as the
physical nature of the accretion disk.

\section{The basics of the broad iron line}

\subsection{Line production}

A substantial amount of the power in AGN is thought to be emitted as
X-rays from the accretion disk corona in active or flaring regions.
Thermal Comptonization (i.e. multiple inverse Compton scattering by hot
thermal electrons; Zdziarski et al. 1994) of soft optical/UV disk
photons by the corona naturally gives rise to a power-law X-ray
spectrum.  The flares irradiate the accretion disk which is relatively
cold resulting in the formation of a ``reflection'' component within the
X-ray spectrum. A similar component is produced in the Solar spectrum by
flares on the solar photosphere (Bai \& Ramaty 1978), in X-ray binaries
by irradiation of the stellar companion (Basko 1978) and in accreting
white dwarfs.

The basic physics of X-ray reflection and iron line fluorescence can
be understood by considering a hard X-ray (power-law) continuum
illuminating a semi-infinite slab of cold gas.  When a hard X-ray
photon enters the slab, it is subject to a number of possible
interactions: Compton scattering by free or bound
electrons\footnote{Whether the electrons are bound or free is of
little consequence for X-rays above 1~keV incident on gas mostly
composed of hydrogen (Vainshtein et al 1983).}, photoelectric
absorption followed by fluorescent line emission, or photoelectric
absorption followed by Auger de-excitation.  A given incident photon
is either destroyed by Auger de-excitation, scattered out of the slab,
or reprocessed into a fluorescent line photon which escapes the slab.

\begin{figure*}[t]
\centerline{\psfig{figure=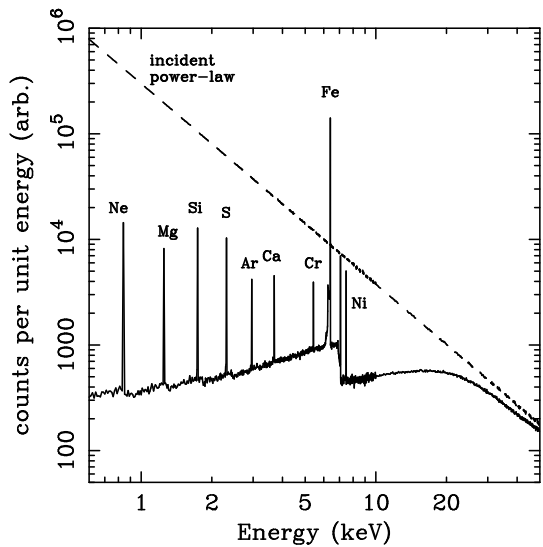,width=0.66\textwidth,angle=270}}
\caption{X-ray reflection from an illuminated slab.  Dashed line shows the
  incident continuum and solid line shows the reflected spectrum
  (integrated over all angles).  Monte Carlo simulation from Reynolds
  (1996).}
\end{figure*}

Figure~1 shows the results of a Monte Carlo calculation which includes
all of the above processes (Reynolds 1996; based on similar
calculations by George \& Fabian 1991).  Due to the energy dependence
of photoelectric absorption, incident soft X-rays are mostly absorbed,
whereas hard photons are rarely absorbed and tend to Compton scatter
back out of the slab. The reflected continuum is therefore a factor of
about $\sigma_{\rm T}/\sigma_{\rm pe}$ below the incident one. Above
energies of several tens of kilovolts, Compton recoil reduces the
backscattered photon flux.  These effects give the reflection spectrum
a broad hump-like shape.  In addition, there is an emission line
spectrum resulting primarily from fluorescent K$\alpha$ lines of the
most abundant metals.  The iron K$\alpha$ line at $6.4\keV$ is the
strongest of these lines.  For most geometries relevant to this
discussion, the observer will see this reflection component superposed
on the direct (power-law) primary continuum.  Under such
circumstances, the main observables of the reflection are a flattening
of the spectrum above approximately 10\,keV (as the reflection hump
starts to emerge) and an iron line at $6.4\keV$.

The fluorescent iron line is produced when one of the
2 K-shell (i.e. $n=1$) electrons of an iron atom (or ion) is ejected
following photoelectric absorption of an X-ray.  The threshold for the
absorption by neutral iron is 7.1~keV. Following the photoelectric
event, the resulting excited state can decay in one of two ways.  An
L-shell ($n=2$) electron can then drop into the K-shell releasing
6.4~keV of energy either as an emission line photon (34 per cent
probability) or an Auger electron (66 per cent probability).  (This
latter case is equivalent to the photon produced by the
$n=2\rightarrow n=1$ transition being internally absorbed by another
electron which is consequently ejected from the ion.) In detail there
are two components to the K$\alpha$ line, K$\alpha_1$ at 6.404 and
K$\alpha_2$ at 6.391~keV, which are not separately distinguished in
our discussion here. There is also a K$\beta$ line at 7.06~keV and a nickel
K$\alpha$ line at 7.5~keV is expected.   

For ionized iron, the outer electrons are less effective at screening
the inner K-shell from the nuclear charge and the energy of both the
photoelectric threshold and the K$\alpha$ line are increased. (The
line energy is only significantly above 6.4~keV when the M-shell is
lost, i.e. FeXVII and higher states.) The fluorescent yield (i.e. the
probability that a photoelectric absorption event is followed by
fluorescent line emission rather than the Auger effect) is also a weak
function of the ionization state from neutral iron (FeI) upto FeXXIII.
For Lithium-like iron (FeXXIV) through to Hydrogen-like iron (FeXXVI),
the lack of at least 2 electrons in the L-shell means that the Auger
effect cannot occur. For He- and H-like iron ions the line is produced
by the capture of free electrons, i.e. recombination. The equivalent
fluorescent yield is high and depends on the conditions (see Matt, Fabian
\& Reynolds 1997).

The fluorescent yield for neutral matter varies as the fourth power of
atomic number $Z^4$, for example being less than one half per cent for
oxygen. Predicted equivalent widths for low $Z$ lines are given in
Matt et al (1997). Fluorescent X-ray spectroscopy is a
well-known, non-invasive way to determine the surface composition of
materials in the laboratory, or even of a planetary surface.

For cosmic abundances the optical depth to bound-free iron absorption is
higher than, but close to, the Thomson depth, The iron line production
in an X-ray irradiated surface therefore takes place in the outer
Thomson depth. This is only a small fraction of the thickness (say 1 to
0.1 per cent) of a typical accretion disk and it is the ionization state
of this thin skin which determines the nature of the iron line.

The strength of the iron line is usually measured in terms of its
equivalent width with respect to the direct emission. (The equivalent
width is the width of the continuum in, say eV, at the position of the
line which contains the same flux as the line. Its determination is
not entirely straightforward when the line is very broad.) It is a
function of the geometry of the accretion disk (primarily the solid
angle subtended by the ``reflecting'' matter as seen by the X-ray
source), the elemental abundances of the reflecting matter, the
inclination angle at which the reflecting surface is viewed, and the
ionization state of the surface layers of the disk.  We will address
the last three of these dependences in turn.

\subsubsection{Elemental abundance}

Elemental abundances effect the equivalent width of the iron line both
through the amount of iron that is present to fluoresce, and the
absorption of the line photons by L-shell photoelectric absorption of
iron and K-shell photoelectric absorption of lower-$Z$ elements. These
competing effects, together with the fact that the edge is saturated
(i.e. most incident photons just above the photoelectric edge are
absorbed by iron ions), leads to a roughly logarithmical dependence on
abundance.  For example, using the cosmic abundance values from Anders
\& Grevesse (1982), the equivalent width as a function of the iron
abundance $A_{\rm Fe}$ is given by
\begin{eqnarray}
W(A_{\rm Fe})=W(A_{\rm Fe}=1)\,(A_{\rm Fe})^\beta & \hspace{2cm}(0.1<A_{\rm Fe}<1)\\
W(A_{\rm Fe})=W(A_{\rm Fe}=1)\,[1+b\log(A_{\rm Fe})] & \hspace{2cm}(1<A_{\rm Fe}<20)
\end{eqnarray}
where
\begin{eqnarray}
(\beta,b)=(0.85,0.95) & \hspace{2cm}{\rm edge-on}\\
(\beta,b)=(0.75,0.48) & \hspace{2cm}{\rm face-on}\\
(\beta,b)=(0.78,0.58) & \hspace{2cm}{\rm angle-averaged}
\end{eqnarray}
where $A_{\rm Fe}=1$ refers to cosmic abundances (Matt, Fabian \&
Reynolds 1997).

\subsubsection{Inclination angle}

As the inclination angle at which the disk is viewed is increased, the
observed equivalent width is depressed due to the extra absorption and
scattering suffered by the iron line photon as it leaves the disk
surface at an oblique angle.  Ghisellini, Haardt \& Matt (1994) find
that $$I(\mu)={{I(\mu=1)}\over{\ln 2}}\mu\log(1+{1\over\mu}),$$ where
$\mu=\cos i$, $i$ being the angle between the line-of-sight and the
normal to the reflecting surface.

\subsubsection{Ionization of the disk surface}

\begin{figure*}[t]
\centerline{\psfig{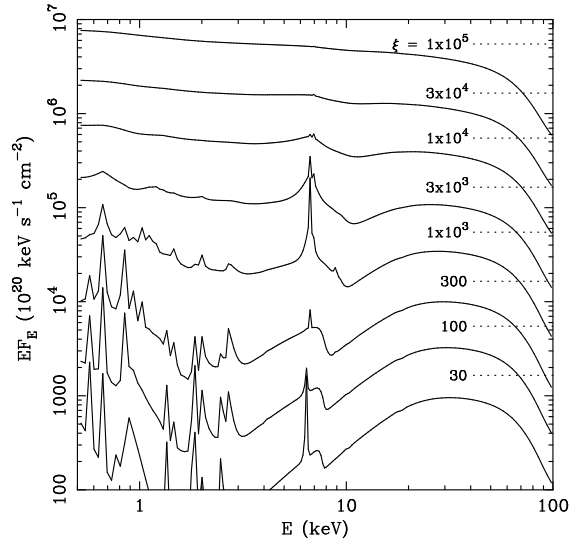}}
\caption{Reflection spectra from ionized matter for various values of
the ionization parameter $\xi$.   The dotted lines show the level of the
illuminating power-law continuum for each value of $\xi$.}
\end{figure*}

X-ray irradiation can photoionize the surface layers of a disk (Ross
\& Fabian 1993; Ross, Fabian \& Young 1999).    As discussed above,
the fluorescent line that the illuminated matter produces depends upon
its ionization state. A useful quantity in this discussion is the
ionization parameter 
$$\xi(r)=4\pi F_{\rm x}(r)/n(r),$$ 
where $F_{\rm x}(r)$ is the X-ray
flux received per unit area of the disc at a radius $r$, and $n(r)$ is
the comoving electron number density: it measures the ratio of the
photoionization rate (which is proportional to $n$) to the
recombination rate (proportional to $n^2$). The iron line emission for
various ionization parameters has been investigated by Matt et al
(1993, 1996).    They found that the behaviour split into four regimes
depending on the value of $\xi$ (also see Fig.~2).

\begin{enumerate}
\renewcommand{\theenumi}{\roman{enumi}}
\item $\xi <100$ ergs cm s$^{-1}$: The material is weakly ionized.
X-ray reflection from the accretion disk produces a cold iron line at
6.4 keV.  Since the total photoelectric opacity of the material is
large even below the iron edge, the Compton backscattered continuum
only weakly contributes to the observed spectrum at 6\,keV, and the
observed iron K-shell absorption edge is small. This regime is termed
`cold' reflection, since the reflection spectrum around the energy of
the iron-K features resembles that from cold, neutral gas.

\item 100 ergs cm s$^{-1}<\xi <500$ ergs cm s$^{-1}$:  In this
intermediate regime, the iron is in the form of FeXVII-FeXXIII and
there is a vacancy is the L-shell ($n=2$) of the ion.  Thus, these
ions can resonantly absorb the corresponding K$\alpha$ line photons.
Successive fluorescent emission followed by resonant absorption
effectively traps the photon in the surafec layers of the disk until
it is terminated by the Auger effect.  Only a few line photons can
escape the disk leading to a very weak iron line.   The reduced
opacity below the iron edge due to ionization of the lower-$Z$
elements leads to a moderate iron absorption edge.
\item 500 ergs cm s$^{-1}<\xi <5000$ ergs cm s$^{-1}$:  In this
regime, the ions are too highly ionized to permit the Auger effect.
While the line photons are still subject to resonant scattering, the
lack of a destruction mechanism ensurs that they can escape the disk
and produces a `hot' iron line at $\sim 6.8\keV$.  There is a large
absorption edge.
\item $\xi >5000$ ergs cm s$^{-1}$: When the disk is highly ionized,
it does not produce an iron line because the iron is completely
ionized. There is no absorption edge.
\end{enumerate}

Note that ionization of the reflector paradoxically causes the
observed iron edge to strengthen at moderate values of $\xi$. This
is because the edge is saturated in reflection from a cold absorber,
as is absorption at lower energies where oxygen and iron-L are the
main absorbers. Ionization of oxygen and iron leads to the iron-K edge
being revealed, and so apparently becoming stronger, as the reflected
flux below the edge increases.

The Matt et al. (1993, 1996) calculations assume a fixed density
structure in the atmosphere of the accretion disk.  Nayakshin, Kallman
\& Kazanas (1999) have relaxed this assumption and included the effect
of thermal instability in the irradiated disk atmosphere.  In their
solutions, the cold dense disk that produces the X-ray reflection
features is blanketed with an overlying low-density, highly ionized,
region.  For weak irradiation, the ionized blanket is thin and does
not affect the observed spectrum.  However, for strong irradiation,
the ionized blanket scatters and smears the ionized reflection
features.  In their models, it can be difficult to produce highly
ionized iron lines in reflection --- the effect of increasing
ionization is to dilute the `cold' reflection signature. The extent of
this effect will depend on the Compton temperature of the radiation
field.

\subsection{The profile of the broad iron line}

\begin{figure}
\centerline{\psfig{figure=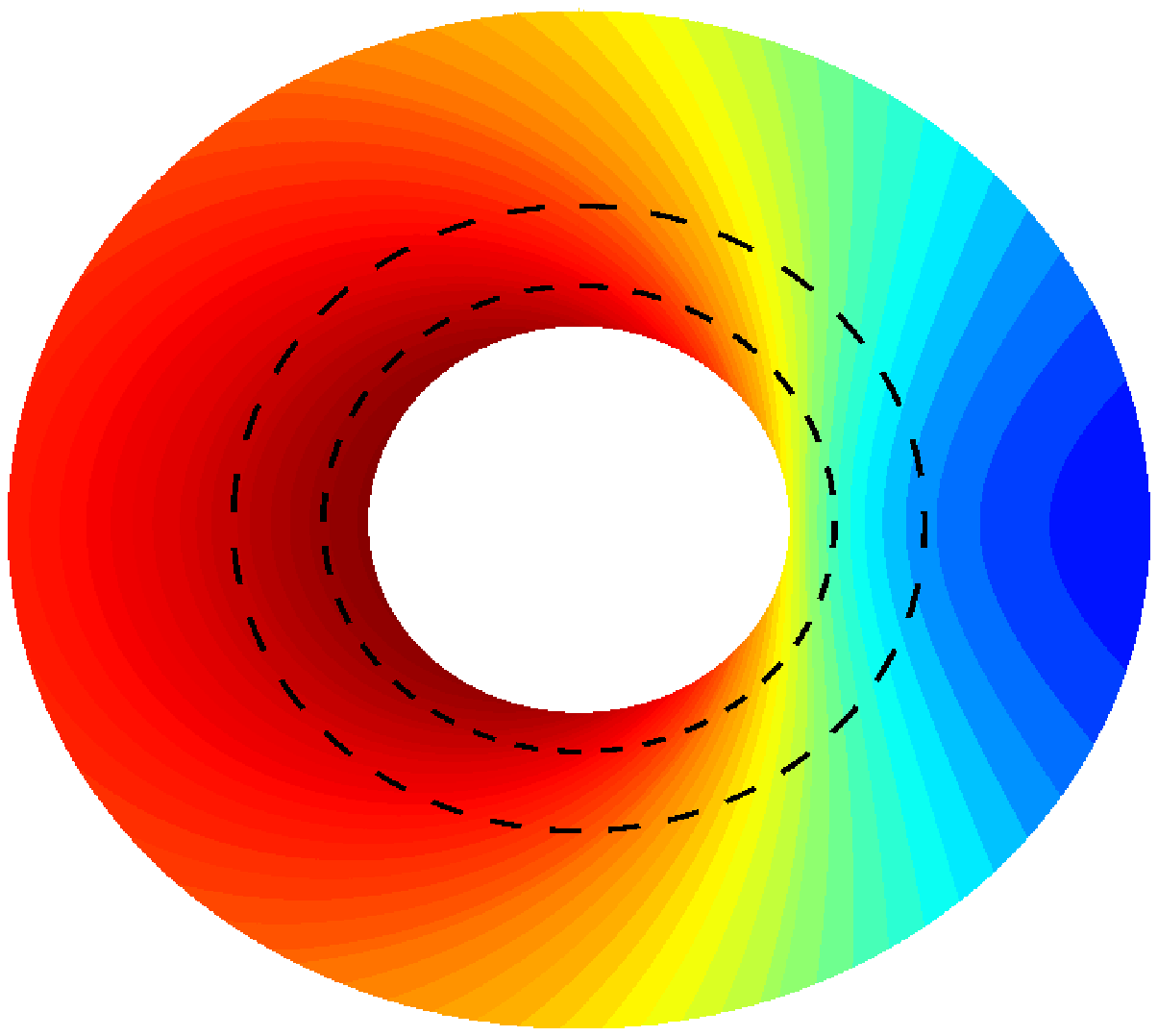,width=0.4\textwidth,angle=0}}
\centerline{\psfig{figure=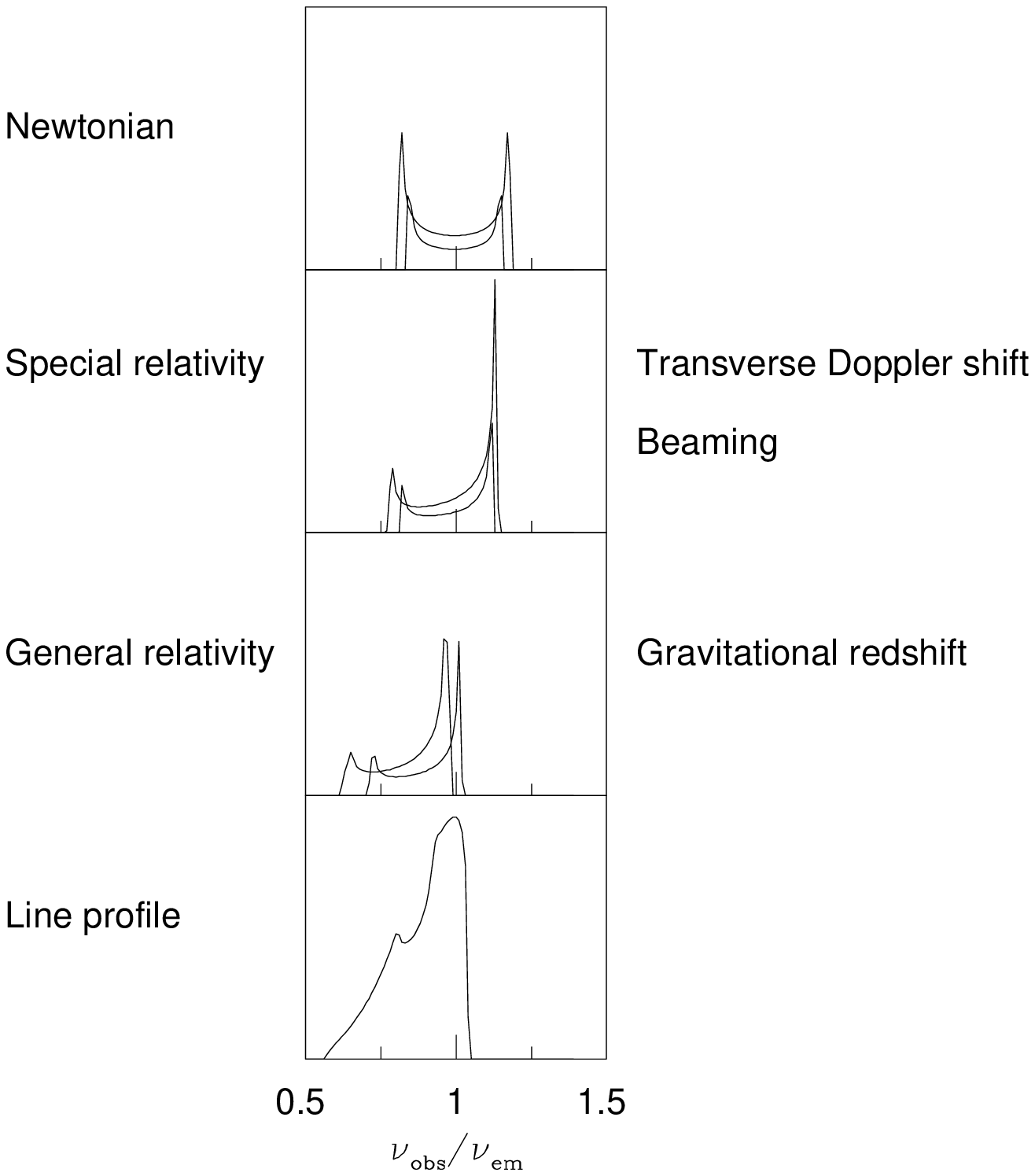,width=0.75\textwidth,angle=0}}
\caption{The profile of the broad iron line is caused by the
  interplay of Doppler and transverse-Doppler shifts, relativistic
  beaming and gravitational redshifting. The upper panel shows the
  symmetric double peaked profiles from two narrow annuli on a
  non-relativistic disk. In the second panel the effects of transverse
  Doppler shifting and relativistic beaming have been included, and in
  the third panel gravitational redshifting has been included. These
  give rise to a broad, skewed line profile, such as that show in the
  lower panel. A more detailed discussion of this figure is given in
  section~2.2.}
\end{figure}

The iron K$\alpha$ line is intrinsically a rather narrow line.  Hence,
we can use broadening of the line to study the dynamics of the
accretion disk.  The line profiles is shaped by the effects of Doppler
shifts and gravitational redshifting.  Figure~3 demonstrates these
effects at work in a schematic way.  In a non-relativistic disk, each
radius of the disk produces a symmetric double-horned line profile
corresponding to emission from material on both the approaching
(blue-shifted) and receding (red-shifted) side.  The inenr regions of
the disk, where the material is moving the fastest, produce the
broadest parts of the line.  Near a black hole, where the orbital
velocities of the disk are mildly relativistic, special relativistic
beaming enhances the blue peak of the line from each radius (second
panel of Fig.~3).  Finally, the comparable influences of the
transverse Doppler effect (i.e. ``moving clocks run slowly'') and
gravitational redshifting (i.e. ``clocks near black holes run
slowly'') shifts the contribution from each radius to a lower energy.
Summing the line emission from all radii of the relativistic disk
gives a skewed and highly broadened line profile.   It has been
suggested by Pariev \& Bromley (1998) that turbulence in the accretion
disk may also significantly broaden the line.   While a detailed
assessment of this possibility must await future magneto-hydrodynamic
disk simulations, it seems unlikely that the turbulent velocity field in
a thin accretion disk will be large enough to broaden the line.

\begin{figure*}[t]
\centerline{\psfig{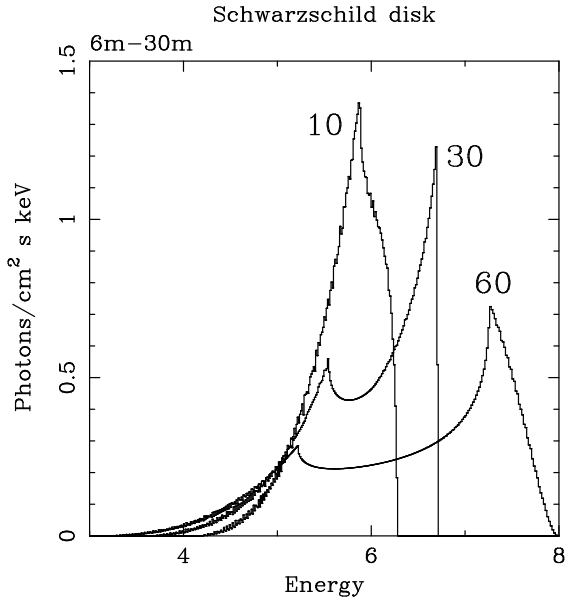}}
\caption{Relativistic iron line profiles for the case of an accretion
disk around a Schwarzschild (non-rotating) black hole.   It is assumed
that the fluorescing region of the disk extends from $6r_{\rm g}$
(i.e. the radius of marginal stability) to $30r_{\rm g}$.   Three
inclinations are shown: $10^\circ$, $30^\circ$ and $60^\circ$.   The
main effect of increasing the inclination is to broaden the line by
increasing its high-energy extent.}
\end{figure*}

\begin{figure*}[t]
\centerline{\psfig{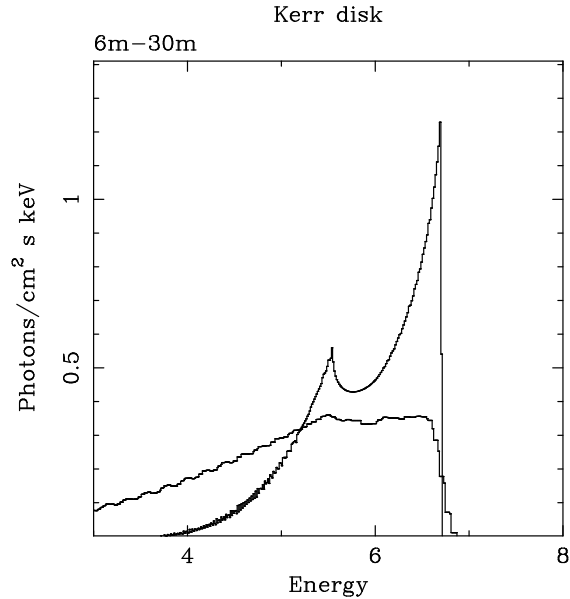}}
\caption{Comparison of relativistic iron line profiles from an
accretion disk around a Schwarzschild black hole (narrower, peaky line)
and a near-extremal Kerr black hole (broader line).  The line emission
is assumed to extend down to the radius of marginal stability which is
$6r_{\rm g}$ and $1.25r_{\rm g}$, respectively.  The difference in the
width and redshift of the line is principally a result of the difference
in the position of the radius of marginal stability.}
\end{figure*}

Some fully relativistic model line profiles are plotted in Figs.~4 and
5. In Fig.~4, we show the line profile from an accretion disk in orbit
around a non-rotating black hole (described by the Schwarzschild
metric).  The line is assumed to be emitted from an annulus of the disk
extending between 6$r_{\rm g}$ and 30$r_{\rm g}$ from the black hole, where
$r_{\rm g}=GM/c^2$ is the standard gravitational radius.  It is seen that the
high energy ``bluewards'' extent of the line is a strong function of the
inclination of the disk.  In fact, the blue extent of the line is almost
entirely a function of the inclination, thereby providing a robust way
to measure the inclination of the disk.  On the other hand, the redward
extent of the line is a sensitive funcion of the inner radius of the
line emitting annulus.  In Fig.~5, we show model iron lines from a
Schwarzschild black hole and a rapidly rotating black hole (described by
a near extremal kerr metric).  In this figure, we have made the
assumption that the line emission extends down to the innermost stable
orbit of the acretion disk.  For these purposes, the principal
difference between these two space-time geometries is the location of
the innermost stable orbit (and hence the inner edge of the line
emission) --- this critical radius is at $6r_g$ in the Schwarzschild
case, and $r_g$ in the extremal kerr case.  

Model line profiles are given in the Schwarzschild case by Fabian et al
(1989) and for the maximal Kerr (spinning black hole) case by Laor
(1991).  Iron lines in extreme kerr metrics are also computed by
Bromley, Miller \& Pariev (1998), and Martocchia, Karas \& Matt (2000).
These last two sets of authors have presented diagnostics that can be
used by observers who wish to avoid full spectral fitting of complex
relativistic models.   The formalism of computing relativistic line
profiles is also discussed by Fanton et al. (1997).

\subsection{Observations of broad iron lines} 

\begin{figure*}[t]
\centerline{\psfig{figure=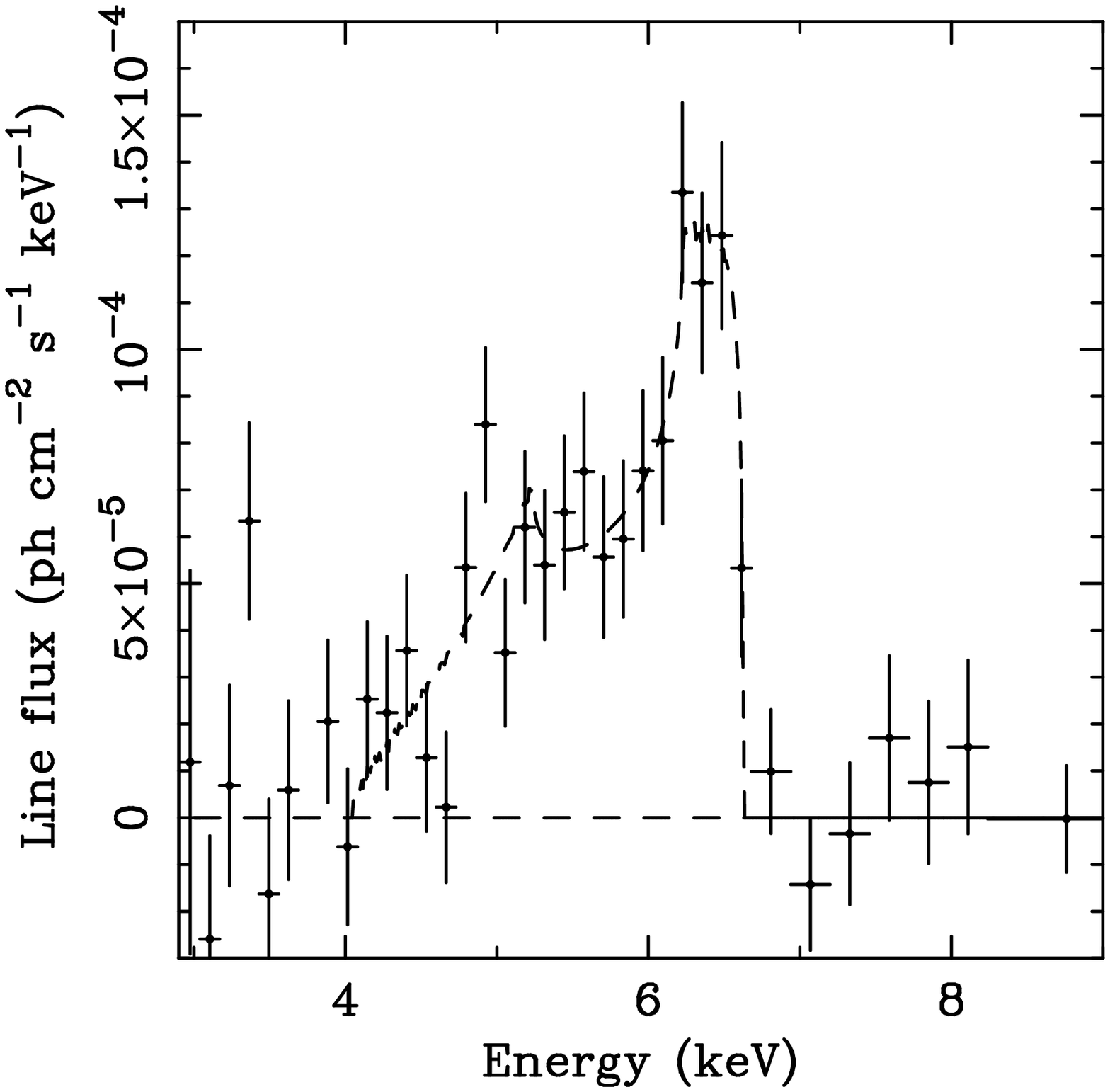,width=0.6\textwidth,angle=270}}
\caption{The line profile of iron K emission in the ASCA SIS
spectrum of the Seyfert 1 galaxy MCG--6-30-15 (Tanaka et al 1995). The
emission line is very broad, with full width at zero intensity of
$\sim 100,000$ km s$^{-1}$. The line shape is skewed toward energies
lower than the rest-energy of the emission line (6.35 keV at the
source redshift of 0.008). The dotted line shows the best-fit line
profile from the model of Fabian et al (1989), an
externally-illuminated accretion disk around a Schwarzschild black
hole.}
\end{figure*}

The X-ray reflection spectrum was first clearly seen by the Japanese
X-ray observatory {\it Ginga} (Pounds et al 1990; Matsuoka et al
1990). The CCD detectors on board the {\it Advanced Satellite for
Cosmology and Astrophysics} ({\it ASCA}) were the first X-ray
spectrometers to provide sufficient spectral resolution and sensitivity
for investigating the profile of the iron line in detail. The first
clear example of a broad skewed iron line came from a long ASCA observation
of the Seyfert 1 galaxy MCG--6-30-15 (Fig. 6; Tanaka et al 1995).  The
sharp drop seen at about 6.5 keV both demonstrates the good spectral
resolution of the CCD detector and, as discussed above, constrains the
inclination of the disk to be about 30 degrees.  If the inclination were
greater then this blue edge to the line moves to higher energies (as
seen in the broad iron line of the Seyfert 2 galaxy IRAS 18325--5926,
Iwasawa et al 1996a). The redward extent of the line constraints the
inner radius of the emission to be 7\rg and the overall shape means that
most of the line emission is peaked within 20\rg. 

Nandra et al. (1997a) and Reynolds (1997) used {\it ASCA} data to
study the iron line in over 20 Seyfert 1 galaxies and found that most are
significantly broader than the instrumental resolution.  In a typical
{\it ASCA} observation of an AGN, the signal-to-noise ratio of the
detected iron line is insufficient to study the line beyond simply
measuring that it is broad.  To combat this problem, Nandra et al.
(1997a) have summed together the data from many AGN to produce an
average iron line profile.  They find that the average line has clear
extension to low energies.  To the extent that individual sources can be
studied, the inferred inclinations of the accretion disks are clustered
around 30 deg, indicating some bias to the selected galaxies.  Such a
bias is expected within the context of the ``unified model'' of Seyfert
galaxies (Antonucci 1993).  In brief, the unified model states that
Seyfert galaxies possess an obscuring torus on scales larger than the
accretion disk.  When one views the central regions of the AGN along a
line of sight that is not blocked by the torus, one seens a type-1
Seyfert galaxy.  Otherwise, one would see a type-2 Seyfert galaxy.  If
the accretion disk and tori are co-aligned (as might be expected on the
basis of dynamical models; Krolik \& Begelman 1988), and the tori have an
average opening angle of 30--40 degrees, then we would naturally expect
a bias in the measured disk inclinations in a sample of Seyfert 1
galaxies.

It should be noted that, in some cases, results on the inclination of
the disk implied from the broad iron line and orientations of the
systems inferred from the other techniques, e.g., ionization cone, radio
jet, broad line clouds etc., differ from each other (e.g., Nishiura,
Murayama \& Taniguchi 1998). The prime example is NGC4151, for which the
iron line suggests the inner accretion disk to be almost face-on (Yaqoob
et al 1996) whilst the other observations point to an edge-on system.
The recent suggestion by Wang et al. (1999) that a significant
proportion of the iron line in NGC4151 is scattered into our
line-of-sight by an electron scattering disk atmosphere may explain this
discrepancy.  Also, such a difference in geometry depending on scales of
interest may be expected due to warping of the accretion disk or a
multiple merger (e.g., NGC1068, Bland-Hawthorn \& Begelman 1997).  Iron
lines from warped accretion disks have been studied theoretically in
some detail by Hartnoll \& Blackman (2000).  In other objects the
inclination inferred from the iron line agrees with the classification
of the object (e.g. MCG--5-23-16; Weaver, Krolik \& Pier 1998).

\begin{figure*}[t]
\centerline{\psfig{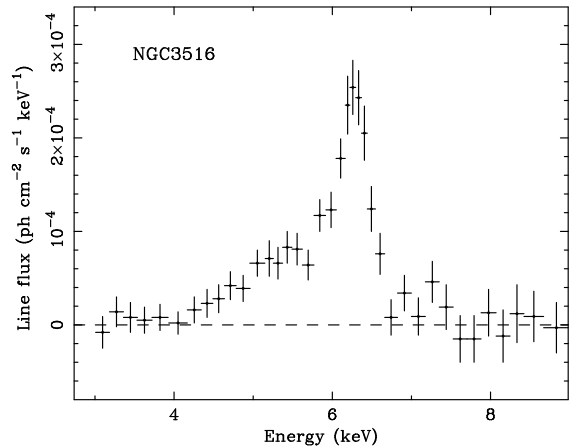}}
\caption{The time-averaged iron line profile observed in the Seyfert
galaxy NGC3516, obtained from a long ASCA observation (Nandra et al
1999). It shows a broad red-tail as well as a resonant absorption
feature around 5.4 keV.}
\end{figure*}

The observed iron line profiles in active galaxies are not necessarily
solely from the accretion disk. Absorption and extra emission may alter
the iron line emitted by the accretion disk. Nandra et al
(1999) reported detection of an absorption feature at about 5.7\,keV
imposed on the broad iron line profile obtained from a long observation
of NGC3516 (see Fig.~7). They suggest that this feature is due to
K$\alpha$ resonant line scattering by highly ionized iron (with an
intrinsic energy of 6.9\,keV).  The redshift of the absorption feature
has been interpreted as evidence for matter infalling onto a black hole.
However, gravitational redshifting of resonant absorption which could
occur when the line photons are passing through the hot corona above the
disk can also account for the observed feature if it occurs close to the
black hole (Ruszkowski \& Fabian 2000).

The iron line profiles observed in some Seyfert galaxies, especially
Compton-thin Seyfert 2s (or those classified as Seyfert 1.8 or 1.9 in
optical), have significant contribution of a narrow line component
originating from matter far away from a the central black hole, e.g., a
molecular torus (Weaver \& Reynolds 1998). Such a narrow component would
become clear if a primary X-ray source had faded away as has been seen
in NGC2992 (Weaver et al 1996) and NGC4051 (Guainazzi et al 1998), since
light travel times puts a fundamental limit on how rapidly line emission
from the torus can respond to the central source.  The iron line
variability observed in NGC7314 (Yaqoob et al 1996) is exactly what is
expected from a line consisting of a broad component originating from
the accretion disk and a torus line.  In this source, the broad iron
line responds to changes in the primary power-law X-ray flux, while a
narrow line component is found to be constant.  It is interesting to
note that this composite nature is not always found in Seyfert 1
galaxies. Any contribution of a stable, narrow line component in
MCG--6-30-15 has been found to be very small (Iwasawa et al 1996; 1999).
This might suggest that, in comparison to Seyfert 2 galaxies, Seyfert 1
galaxies have tori that have either smaller optical depths or smaller
geometric covering factors.

\begin{figure*}[t]
\centerline{\psfig{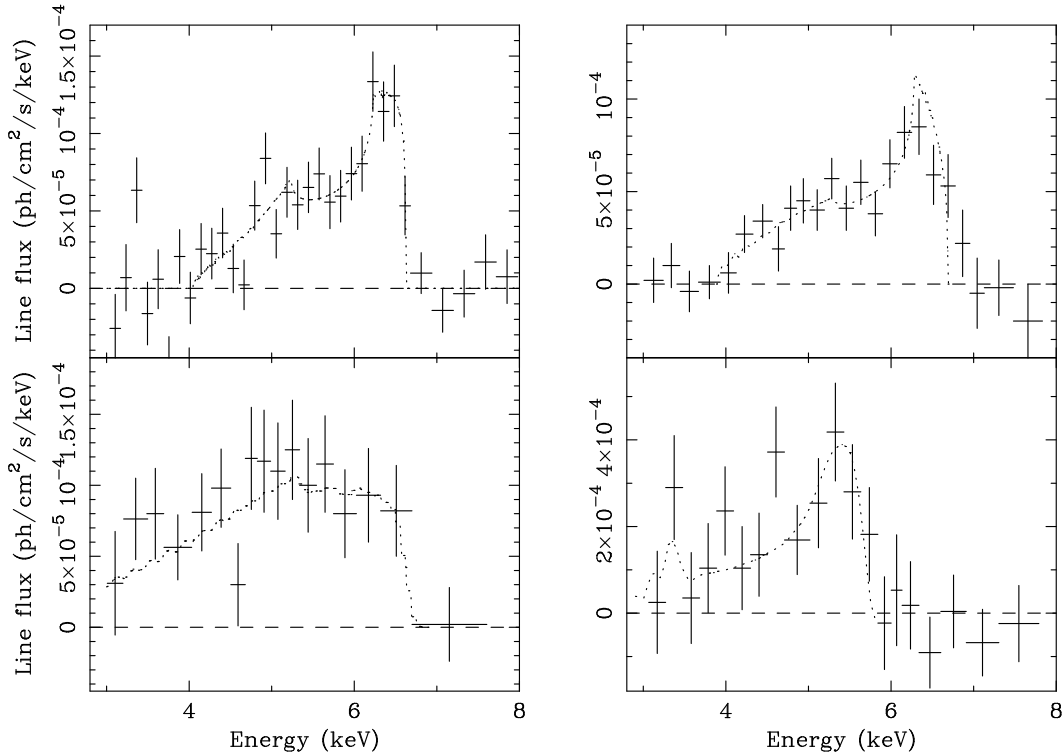}}
\caption{Time-averaged (upper panels) and peculiar line profiles
(lower panels) of the iron K emission from MCG--6-30-15 seen in the
two long ASCA observations in 1994 (left) and 1997 (right). In the
1994 observation, a very broad profile with a pronounced red-wing is
seen during a period of Deep Minimum of the light curve (lower left),
compared to the time-averaged line profile shown in the upper
panel. In contrast, during a sharp flare in the 1997 observation,
whole line emission is shifted to energies below 6 keV and there is no
significant emission at the rest line-energy of 6.4 keV (lower right).
Both peculiar line shapes can be explained by large gravitational
redshift in small radii on the accretion disk.}
\end{figure*}

Not all bright Seyfert 1 galaxies have iron lines so broad that the
disk is required to extend down to the marginally stable orbit (i.e.
$<6r_{\rm g}$). IC4329a is found to have a relatively narrow line
(Done et al 2000) which can be modelled with a disk of inner radius
of about $50r_{\rm g}$. The disk within this radius may either be
missing or highly ionized.

Rapid X-ray variability in active galaxies predicts the iron line also
to vary in response to the continuum with a small time lag. The light
crossing time for 10\rg in a black hole with a mass of $10^7M_7\Msun$ is
about 500$r_1M_7$ s, which is much shorter than an integration time
required for {\it ASCA} to collect enough line photons to perform a
meaningful measurement in X-ray bright AGN ($>10^{-11}$
\ergpcmsqps). It means that no reverberation effects in the line
can be detected with ASCA.  We will return to reverberation effects in
Section 5.  

Despite the inability of current detectors to measure reverberation,
significant and complex variations of the iron line in profile as well
as intensity have been observed in the Seyfert 1 galaxy MCG--6-30-15
(see Fig.~8; Iwasawa et al 1996b).  During the first long observation of
this galaxy in 1994, a line profile with unusually strong blue peak was
found in a time interval of a bright flare whilst the line showed a
very-broad, red-wing dominated profile during a deep minimum period.  In
this very-broad state, line emission from within 6\rg\ is required to
explain the line profile and width.  Possible theoretical
interpretations of this are discussed in Section 4.  A succession of
large flares on the approaching side of the disk could produce the
blue-peak dominated line profile, although it can also be explained if
the line is produced predominantly at large radii ($\sim 100$\rg). It is
worth noting that this bright flare showed a continuum spectral
evolution similar to that seen in a shot in a Galactic black hole
candidate, e.g., Cyg X-1 (Negoro et. al 1995).

Another peculiar line shape seen in a brief period ($\sim 1$ hr) of a
flare during the 1997 long observation of MCG--6-30-15 also requires a
large redshift within 6 r$_{\rm g}$. The blue wing of this line is, this
time, shifted well below 6 keV and no significant line was detected
around 6.4 keV. A possible explanation is that the line production
occurred either in a thin annulus at 4 r$_{\rm g}$ or a small patch at
$\sim $2.5 r$_{\rm g}$ on the approaching side of the disk (Iwasawa et al
1999).

Detailed studies of the {\it RXTE} data on the iron line variability in
MCG--6-30-15 have presented a puzzling problem; most of the line flux
appears to be constant in spite of strong continuum changes (Lee et al
1999; Reynolds 2000: see also studies of NGC5548 by Chiang et al 1999
and of the Galactic Black Hole Candidate Cygnus X-1 by Revnivtsev,
Gilfanov \& Churazov 1999). There should therefore be some
self-regulating mechanism to produce a constant line flux, which has yet
to be understood. However, a separate investigation of the broad
red-wing and narrow core of the iron line for the 1994 observation
(Iwasawa et al 1996) has revealed interesting behaviours of each
component. The narrow core remains constant on time scales shorter than
$10^3$ s but follows the continuum variations on longer time scales
($>10^4$ s). In contrast, the broad red-wing appears to follow the
continuum on the short time scales. This is consistent with a line
produced from a relativistic disk (see also Blackman 1999).  Also
puzzling is an anti-correlation between the reflected fraction and the
equivalent width of the iron line measured in MCG--6-30-15 (Lee et al
2000) and NGC5548 (Chiang et al 2000).

\section{Alternative models for a broad line} 

The claim that iron line studies are probing the region within a few
gravitational radii of the black hole is a bold one, and should be tested
against other models at every opportunity.  Furthermore, the internal
consistencies of the accretion disk hypothesis must be critically examined.
Given the quality of data, the July-1994 MCG$-$6-30-15 line profile has
become a testbed for such comparisons.

Fabian et al. (1995) examined many alternative models including lines
from mildly relativistic outflows, the effect of absorption edges on the
observed spectrum, and broadening of the line via comptonization.
Fabian et al. found that none of these models were viable alternatives
for the MCG$-$6-30-15 line profile.  The idea of producing the broad
line via Comptonization has been revived recently by Misra \& Kembhavi
(1997) and Misra \& Sutaria (1999).  They argue that the spectrum
initially consists of a narrow iron line superposed on a power-law
continuum and that Comptonization in a surrounding cloud with optical
depth $\tau\sim 4$ produces the broad line.  The Comptonizing cloud
must be both cold ($kT<0.5\keV$ in order to predominately
downscatter rather than upscatter the line photons), and fully-ionized
(since no strong iron absorption edges are seen in the continuum
spectrum).  The cloud is kept fully ionized and yet cool by postulating
that the continuum source has a very luminous optical/UV component.

There are strong arguments against such a model.  Since the power-law
continuum emission also passes through any such Comptonizing cloud, one
would observe a break in the continuum spectrum at $E_{\rm br}\sim
m_{\rm e}c^2/\tau^2\sim 30\keV$.  Such a break is not observed in the
{\it BeppoSAX} (Guainazzi et al. 1999) or {\it RXTE} data (Lee et
al. 1999) for MCG--6-30-15 (see Misra 1999).  Also, both continuum
variability (which is seen on timescales as short as 100\,s) and
ionization arguments limit the size of the Comptonizing cloud in
MCG--6-30-15 to $R<10^{12}\cm$.  The essence of this ionization argument
is that the ionization parameter at the outer edge of the cloud (which,
for a fixed cloud optical depth, scales with cloud size as $\xi\propto
1/R$) must be sufficiently high that all abundant metals, including
iron, are fully ionized (Fabian et al. 1995; Reynolds \& Wilms 2000).
In the case of MCG--6-30-15, these constraints on the cloud size turn
out to so tight that the postulated optical/UV component required to
Compton cool the cloud would violate the black body limit (Reynolds \&
Wilms 2000). Comptonization moreover provides a poor fit (Ruszkowski
\& Fabian 2000).  Hence, we consider the Comptonization model for the broad
iron line not to be viable.

In another alternative model, Skibo (1997) has proposed that energetic
protons transform iron in the surface of the disk into chromium and
lower $Z$ metals via spallation which then enhances their fluorescent
emission (see Fig.~1).  With limited spectral resolution, such a line
blend might appear as a broad skewed iron line.  This model suffers both
theoretical and observational difficulties.  On the theoretical side,
high-energy protons have to be produced and slam into the inner
accretion disk with a very high efficiency (Skibo assumes $\eta=0.1$ for
this process alone).  On the observational side, it should be noted that
the broad line in MCG--6-30-15 (Tanaka et al 1995) is well resolved by
the {\it ASCA} SIS (the instrumental resolution is about 150~eV at these
energies) and it would be obvious if it were due to several separate and
well-spaced lines spread over 2~keV. There can of course be
doppler-blurring of all the lines, as suggested by Skibo (1997), but it
will still be considerable and require that the redward tails be at
least 1 keV long.

Finally it is worth noting that the line profile indicates that most of
the Doppler shifts are due to matter orbiting at about 30 degrees to the
line of sight. The lack of any large blue shifted component rules out
most models in which the broad line results from iron line emission from
bipolar outflows or jets. What we cannot determine at present is the
geometry in more detail. For example, we cannot rule out a `blobby' disk
(Nandra \& George 1994). We do, however, require that any corona be
either optically-thin or localized, in order that passage of the
reflection component back through the corona does not smear out the
sharp features. (Note though that an optically-thick corona over the
inner regions of a disc would explain the lack of an iron line from that
region.)

\section{What happens inside 6 r$_{\rm g}$: Kerr black holes} 

Circular particle orbits around a black hole are only stable outside the
radius of marginal stability, $r_{\rm ms}$. Within this radius, it is
normally assumed that the material plunges ballistically into the black
hole. The location of $r_{\rm ms}$ depends on the black hole spin,
decreasing from 6 r$_{\rm g}$ for a static Schwarzschild black hole
($a/m=0$) to 1.235 r$_{\rm g}$ for prograde orbits around a maximally
spinning Kerr black hole ($a/m=0.998$). Thus the inner edge of a
Keplerian accretion disk will be at $r_{\rm ms}$. In the standard model
the X-ray emission from the accretion disk takes place in a (possibly
patchy) disk-hugging corona above the (almost) Keplerian part of the
disk. The material inflowing within $r_{\rm ms}$ cannot support a corona
and is assumed to receive an insignificant fraction of the X-ray
illumination. In such models fluorescent line emission is expected to
extend in only as far as $r_{\rm ms}$.

\begin{figure*}[t]
\hbox{
\hspace{-1cm}
\psfig{figure=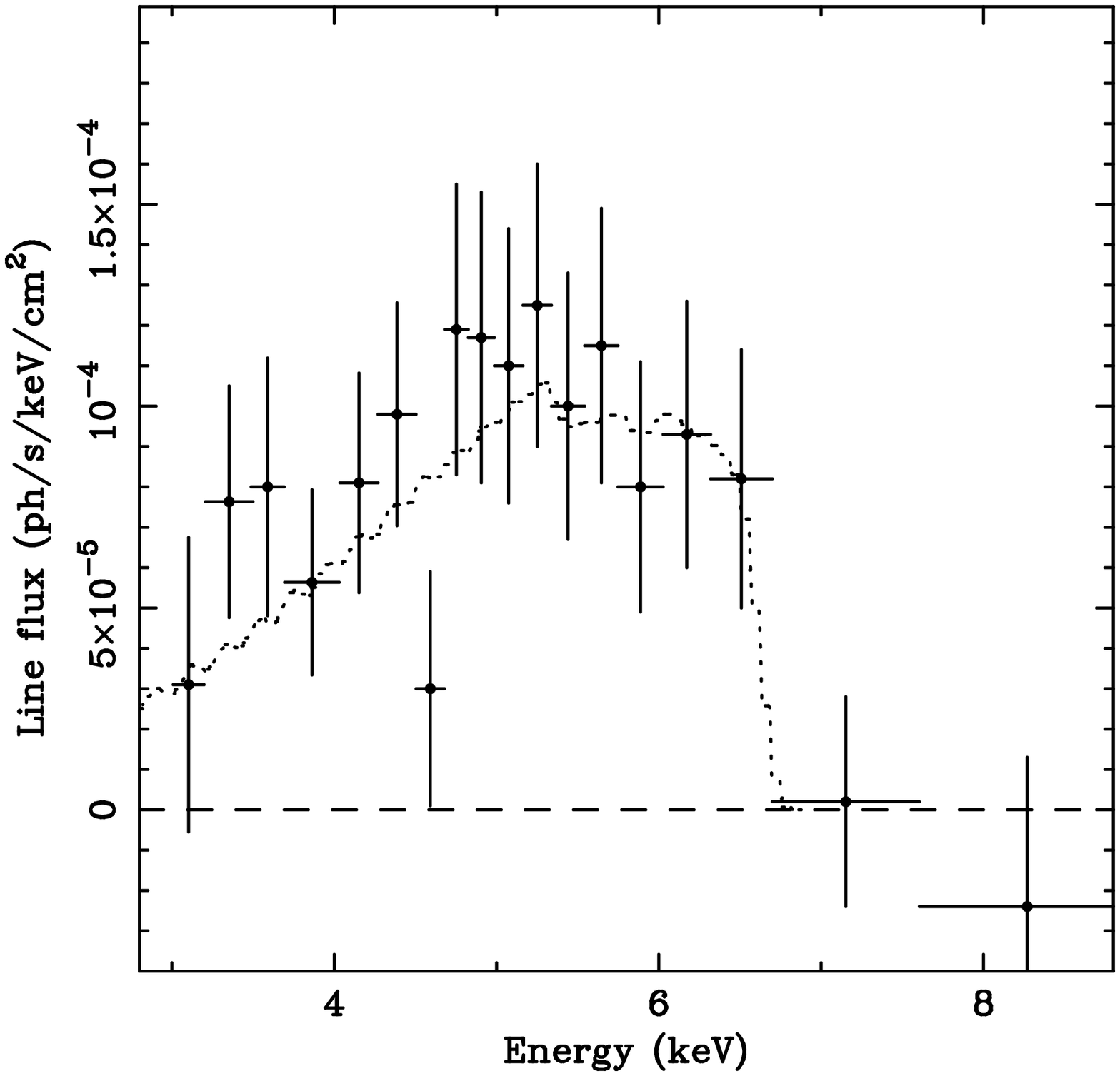,width=0.65\textwidth,angle=270}
\hspace{-2cm}
\psfig{figure=f9b.ps,width=0.45\textwidth,angle=270}
}
\caption{The broad and highly redshifted iron line
profile of MCG--6-30-15 observed during a deep minimum in its light
curve. The left panel shows the line profile from a maximally spinning
Kerr black hole (dotted line; see also the lower left panel of
Fig.~8), and the right panel shows the line profile from a static
Schwarzschild black hole using the model of Reynolds \& Begelman
(1997).}
\end{figure*}

As discussed previously the degree of redshifting seen in the iron line
is an indication of how close the line emitting region extends to the
black hole. During a deep minimum in the X-ray light curve of the 1994
\emph{ASCA} observation of MCG--6-30-15 the iron line was seen to
broaden and shift to lower energies (see Fig. 8). The only way
to have such significantly redshifted line flux is for the source of
emission to move within the innermost stable orbit for a static
Schwarzschild black hole (i.e. 6 r$_{\rm g}$). The line is well fit by
the profile of a maximally spinning Kerr black hole leading to the
tentative conclusion that the line was the first spectroscopic
evidence for a Kerr black hole (Iwasawa et al 1996). (It was tentative
because of the difficulty in measuring the continuum precisely at that
time due to an increase in the strength of the warm absorber; Otani et
al 1996). Later work by Dabrowski et al (1997) quantified the spin of
the black hole required to produce such highly redshifted emission to
exceed 95 per cent of its maximal value.  The equivalent width of the
line also increased dramatically (by a factor of 3--4) during this
deep-minimum.

Of course, the black hole spin cannot change on such short timescales.
It seems, instead, that the pattern of X-ray illumination across the
disk must change in the sense that it becomes more concentrated that
average during the deep minimum.  Since such a dramatic change in
illumination pattern is unlikely to occur just by chance, some
structural change in the X-ray emitting corona is inferred.  A plausible
correspondence of the accretion disk thermal timescale, and the
timescale on which the source enters the deep-minimum, may suggest that
such structural changes are mediated by thermal instabilities in the
accretion disk or accretion disk corona.  An unsolved problem of this
model is that one would expect the continuum level to increase
significantly (rather than suffer the observed decrease) when the
emission is originating from the energetically-dominant, innermost
regions of the accretion disk.

\begin{figure*}[t]
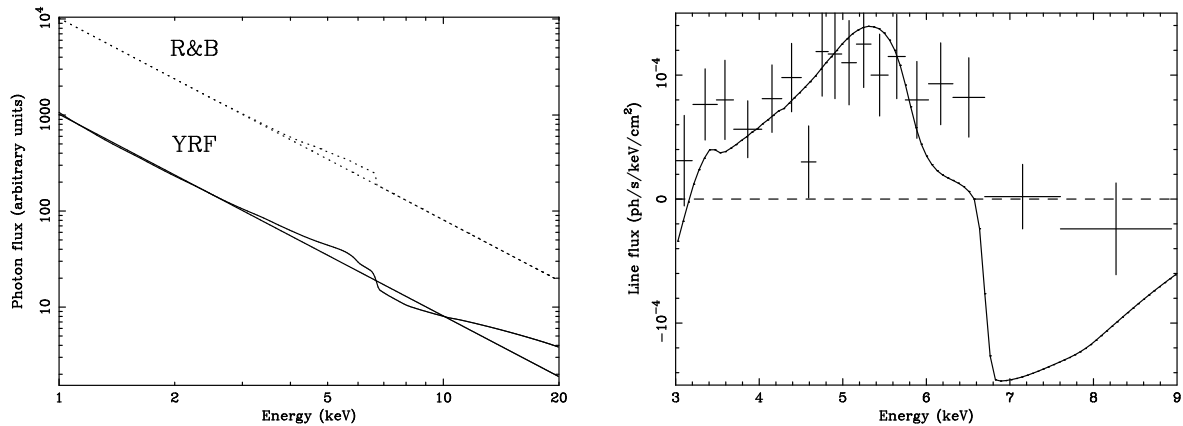

\hbox{
\psfig{figure=f10a.ps,width=0.45\textwidth,angle=270}
\hspace{0.5cm}
\psfig{figure=f10b.ps,width=0.45\textwidth,angle=270}
}
\caption{The modeling of Reynolds \& Begelman (1997) does not take into account
the Compton reflected continuum that accompanies the line. The left
panel shows a comparison of their model (R\&B) with one in which the
full reflected continuum has been taken into account (YRF; Young, Ross
\&Fabian 1998). The reflected continuum should posses a strong
absorption edge that tentatively appears to be inconsistent with the
data (right panel), although future observations are required to
unambiguously determine the spin of the black hole.}
\end{figure*}

Reynolds \& Begelman (1997) have pointed out that although the inner
edge of the accretion disk around a Schwarzschild (i.e. non-rotating)
black hole is at 6 r$_{\rm g}$  the accretion flow does not immediately become
optically thin at smaller radii, and may remain optically thick almost
all the way down to the horizon. If this it is illuminated in the right
manner, ionized iron fluorescence from this region can give rise to an
extremely broad and highly redshifted iron line very similar to that
observed during the deep minimum (see Fig. 9).  In this model, the illuminating
source is assumed to be a point source located on the symmetry axis of
the accretion disk.  The change in line profile, line equivalent width,
and continuum level are explained by assuming that the X-ray source
changes its height above the accretion disk with approximately constant
luminosity.  However, the modelling of Reynolds \& Begelman (1997) did
not account for the Compton reflected continuum which would be expected
to accompany the line.  Young, Ross \& Fabian (1998) have computed both
the line and continuum spectrum expected from such a model. Within $6m$
the density of the infalling material drops rapidly and it becomes
photoionized by the X-ray illumination, generating a large iron
absorption edge (see Fig.~10).  Preliminary indications suggest that
such an edge may not be present in the data, although detailed spectral
fitting has yet to be performed in order to address this issue.  While
the situation with current data remains ambiguous, searching for these
type of spectral features in future {\it XMM--Newton} data may allow us to
unambiguously determine the spin of the black hole in MCG--6-30-15 and
other objects.

The detailed nature of the accretion flow within 6 r$_{\rm g}$ of a
Schwarzschild black hole is likely to be a lot more complicated than we
have assumed in the above discussion. Magnetic fields within the
innermost stable orbit may be rapidly amplified until their energy
density is comparable to the rest-mass energy density of the accretion
flow (Krolik 1999), and hence will be dynamically significant. Such
strong field enhancement within the radius of marginal stability may
lead to the creation of an ``inner'' X-ray emitting corona (note that
Lee et al. 2000 have argued for an inner, highly variable, X-ray corona
on the basis of the observed {\it RXTE} variability). The presence of
magnetic fields may also exert a torque on the inner edge of the disk
(usually assumed to have a zero-stress boundary condition) thereby
increasing the outward flow of angular momentum and the efficiency of
the disk (Agol \& Krolik 2000; although also see Paczy\'nski 2000 for an
opposing view). If there is appreciable line fluorescence from within
the radius of marginal stability, reverberation mapping (discussed
below) may allow this region to be mapped out in detail.

\section{Reverberation} 

The rapid X-ray variability of many Seyfert galaxies leads us to believe
that the primary X-rays are emitted during dramatic flare-like events in
the accretion disk corona.  When a new flare becomes active, the hard
X-rays from the flare will propagate down to the cold disk and excite iron
fluorescence.  Due to the finite speed of light, the illumination from the
flare sweeps across the disk, and the reflected X-rays act as an `echo' of
this flare.  Such flaring will cause temporal changes in the iron line
profile and strength due to the changing illumination pattern of the disk
and, more interestingly, time delays between the observed flare and the its
fluorescent echo.  This latter effect is known as reverberation.

In principal, reverberation provides powerful diagnostics of the
space-time geometry and the geometry of the X-ray source.  When
attempting to understand reverberation, the basic unit to consider is
the point-source transfer function, which gives the response of the
observed iron line to an X-ray flash at a given location.  As a
starting point, one could imagine studying the brightest flares in
real AGN and comparing the line variability to these point-source
transfer functions in an attempt to measure the black hole mass, spin
and the location of the X-ray flare.  By studying such transfer
functions, it is found that rapidly rotating black holes possess a
characteristic reverberation signature.  In the case where the
fluorescing part of the accretion disk extends down to the radius of
marginal stability of a near-extremal kerr black hole, the
instantaneous iron line profile displays a low-energy bump which moves
to lower-energies on a timescale of $GM/c^3$.  This feature
corresponds to highly redshifted and delayed line emission that
originates from an inwardly moving ring of illumination/fluorescence
that asymptotically freezes at the horizon (see Reynolds et al. 1999
and Young \& Reynolds 2000 for a detailed discussion of this feature).

\begin{figure*}[t]
\hbox{
\psfig{figure=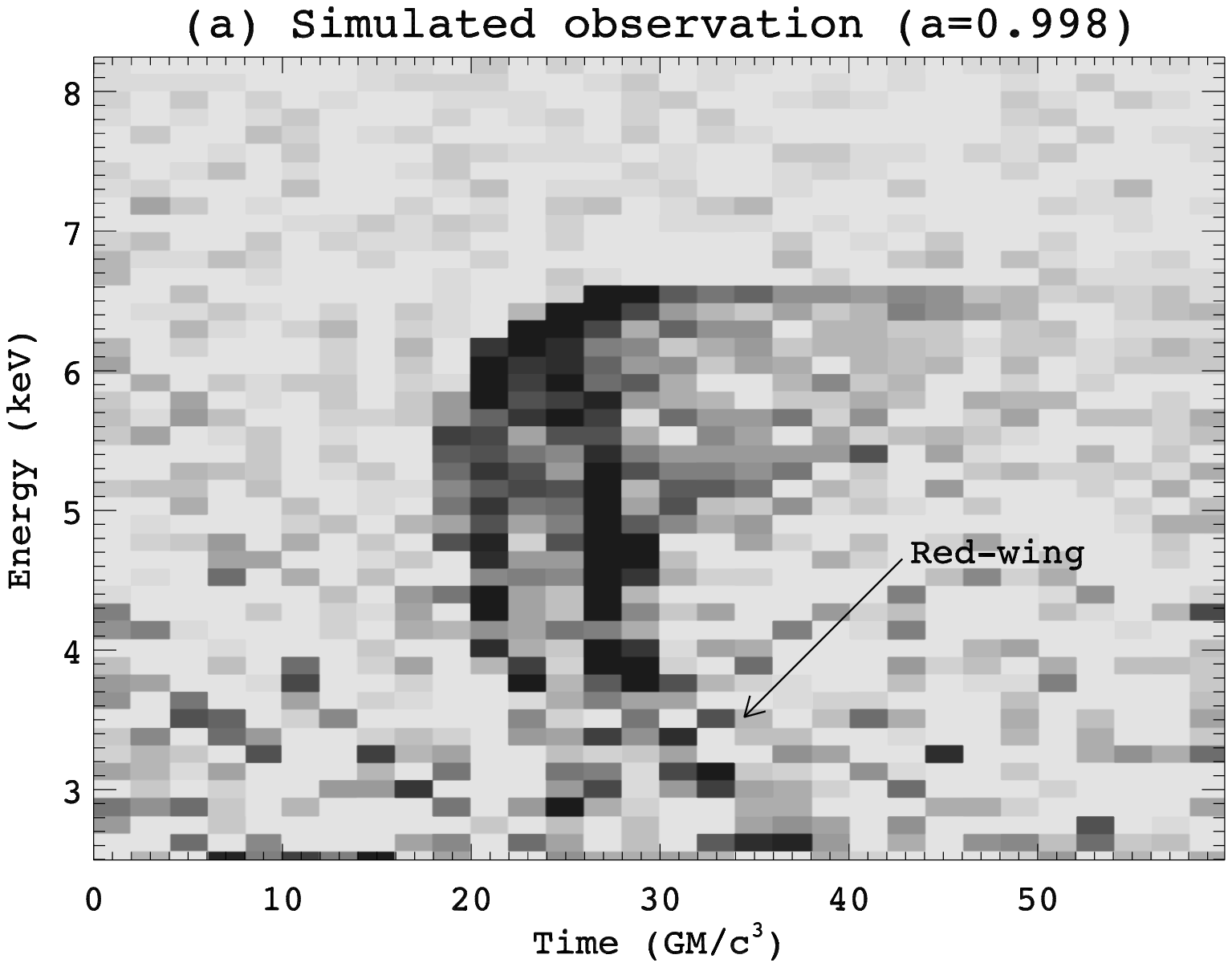,width=0.45\textwidth}
\hspace{0.5cm}
\psfig{figure=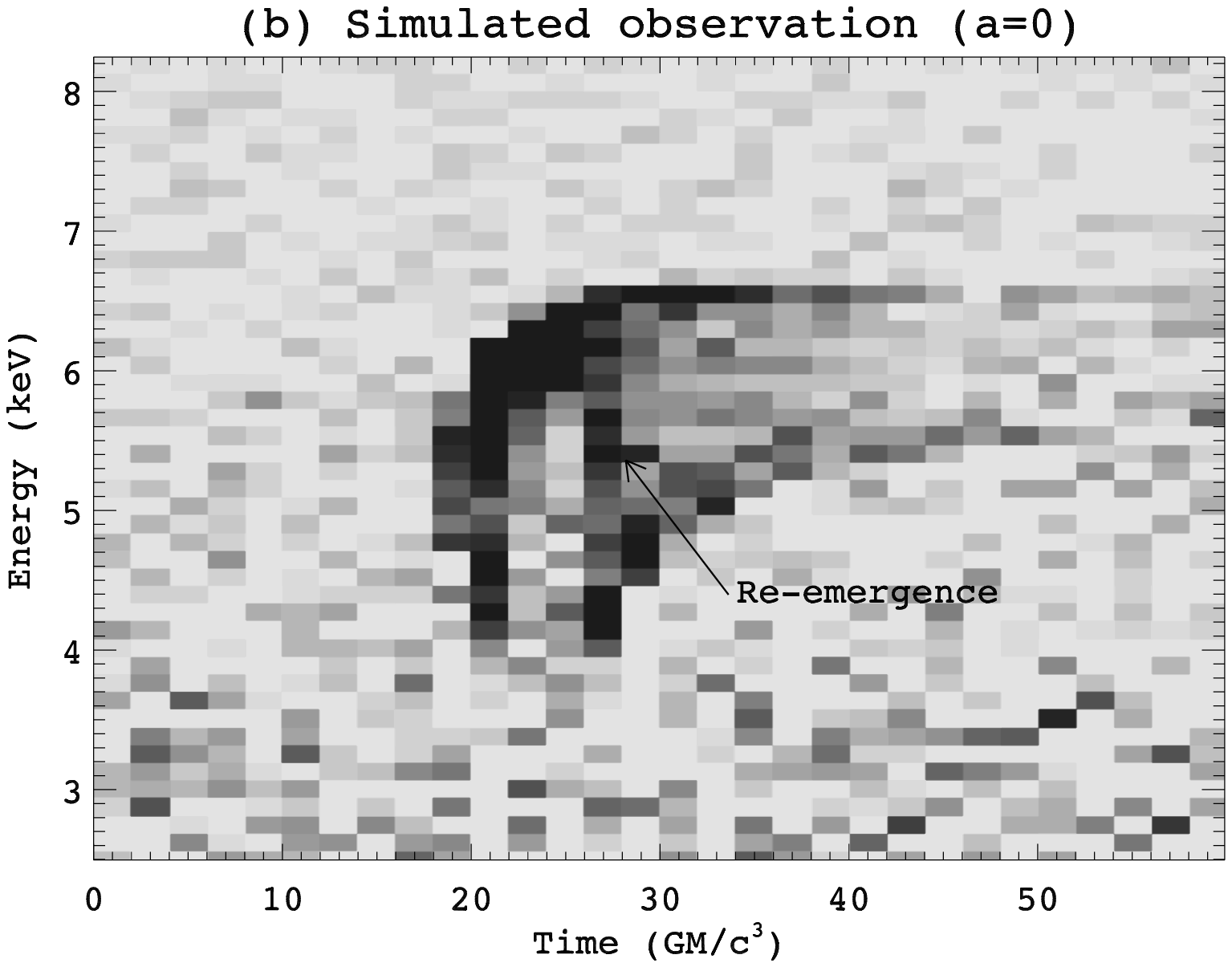,width=0.45\textwidth}
}
\caption{{\it Constellation-X} simulations of iron line reverberation.
Panel (a) shows the case of a rapidly rotating black hole whereas
panel (b) shows a non-rotating black hole.  In both cases, an X-ray
flash on axis at a height of $10GM/c^2$ has been assumed and the iron
line response calculated for an accretion disk inclination (away from
normal) of 30$^\circ$.  Sequential 1000\,s {\it Constellation-X}
observations of the time varying iron line are then simulated,
continuum subtracted, and stacked in order to make an observed
transfer function.  Figure from Young
\& Reynolds (2000).}
\end{figure*}

The primary observational difficulty in characterizing iron line
reverberation will be to obtain the required signal-to-noise.  One must
be able to measure an iron line profile on a timescale of $t_{\rm
reverb}\sim GM/c^3\approx 500 M_8\,{\rm s}$, where we have normalized to
a central black hole mass of $10^8\,{\rm M}_\odot$.  This requires an
instrument with at least the collecting area of {\it XMM--Newton}, and probably
{\it Constellation-X}.  Figure~11 shows that {\it Constellation-X} can
indeed detect reverberation from a bright AGN with a mass of $10^8\,{\rm
M}_\odot$.  Furthermore, the signatures of black hole spin may well be
within reach of {\it Constellation-X} (Young \& Reynolds 2000).
Although these simulations make the somewhat artificial assumption that
the X-ray flare is instantaneous and located on the axis of the system,
it provides encouragement that reverberation signatures may be
observable in the foreseeable future.

Of course, the occurrence of multiple, overlapping flares will also
hamper the interpretation of iron line reverberation.  In the
immediate future, the most promising (although observationally
expensive) approach will be to observe a bright source while it
undergoes a very large flare. (Although the current data show little
evidence in general for correlated behaviour of the continuum and iron
line, as mentioned in Section 2.3, changes in the line are seen during
some of the largest flares and dips in MCG--6-30-15.) Assuming that
the single flare temporarily dominates the flux from the source, we
might expect the point source transfer function to be an adequate
description of the reverberation.  Ultimately, reconstruction
techniques for extracting reverberation signatures from overlapping
flares should be explored.

\section{Other types of black hole systems} 

As outlined in the preceding discussion, most of the progress in the
field of broad iron line observations has been for type-1 Seyfert
galaxies.   In this section, we briefly discuss X-ray reflection
diagnostics of other types of black hole systems.

\subsection{Quasars}

There is no evidence of an iron line or any reflection features in the
spectra of most luminous quasars (Nandra et al 1995; 1997). The
equivalent width of the iron line is observed to decrease with
increasing luminosity as one moves from the Seyfert 1 regime ($L_{\rm
X}\approxlt 10^{44}\ergps$) to the quasar regime, a phenomenon termed
the `X-ray Baldwin effect' (Iwasawa \& Taniguchi 1993; Nandra et al
1995). It has been suggested that the accretion disk becomes
increasingly ionized.  This might be due to the most luminous objects
possessing accretion rates closer to the Eddington limit. One puzzling
aspect of this explanation is that the disk must jump from being `cold'
to being completely ionized otherwise we would observe instances of
intermediate ionization in which the iron in the surface layers of the
disk is H or He-like and the equivalent width of the line is even larger
(Fig.~2 with $\xi=1\times10^3$). There should also be a large absorption
edge that is not seen. As mentioned in Section 2.1, a thermal
instability in the surface layers of the disk (Nayakshin et al 1999) may
lead to the formation of a highly ionized blanket and circumvent this
problem.

The absence of reflection features can also be explained if the
fluorescing accretion disk subtends less than $2\pi\sr$ as seen from the
illuminating X-ray source, e.g. if the fluorescing accretion disk
truncates at a few tens of Schwarzschild radii.  A transition to an
advection dominated accretion flow (ADAF), which is hot and
optically-thin, would produce such a geometry.  However, such structures
can only exist at small acretion rates.  Together with the fact that
most of the energy in such a structure is advected into the black hole,
ADAFs are expected to be {\it much} less luminous than most quasars.  A
more likely possibility is that, as an object approaches its Eddington
limit, the region of the disk that is radiation pressure dominated moves
further out, causing the surface layers of the disk to become more
tenuous and highly ionized.  The formation of a super-Eddington
accretion disk in which much of the radiative energy is trapped in the
accretion flow may also be relevant to quasars.

Curiously, one of the most luminous low redshift quasars, PDS\,456,
shows significant features at the iron-K energies (Reeves et al 2000).
The features appear as a deep ionized edge and a possible broad line,
and are modelled as either an ionized reflector (disk), or less likely
as a strong highly ionized warm absorber.

\subsection{Low luminosity AGN}

Most massive black holes in the local universe are accreting at rates
that are much lower than those found in AGN.  If the accretion is so
small as to render the black hole undetectable, the galaxy is termed a
quiescent galaxy.  Slightly higher accretion rates will lead to the
classification as a {\it low luminosity} AGN (LLAGN).  

The nature of black hole accretion when the accretion rates are very low
is a topic of active research.  It was realized by several authors that
when the accretion rate is low (relative to the Eddington rate), an
accretion disk may switch into a hot, radiatively-inefficient mode
(Ichimaru 1977; Rees 1982; Narayan \& Yi 1994; Narayan \& Yi 1995).  In
essence, the plasma becomes so tenuous that the timescale for energy
transfer from the protons to the electrons (via Coulomb interactions)
becomes longer than the inflow timescale.  The energy remains as thermal
energy in the protons (which are very poor radiators) and gets advected
through the event horizon of the black hole.  These are the so-called
Advection Dominated Accretion Flows (ADAFs).  ADAFs are to be contrasted
with `standard' radiatively-efficient accretion disks in which the disk
remains cool and geometrically thin all of the way down to the black
hole (Shakura \& Sunyaev 1973; Novikov \& Thorne 1974).  Broad iron line
studies of LLAGN provide a potentially important probe of the physics of
accretion when the accretion rate is low --- the iron line traces only
the radiatively-efficient portions of the disk since ADAFs are far too
hot to produce fluorescent iron line emission.

Observationally, LLAGN have proven difficult to study due to the fact
that they are X-ray faint.  In addition, their X-ray spectra are
typically complex with non-nuclear spectral components (such as
starburst regions and/or thermal emission from hot gas) rivalling the
nuclear component (e.g. see Ptak 1997).  One of the best studied LLAGN
resides in the nearby galaxy NGC~4258 (M~106).  A short {\it ASCA}
observation of this galaxy hinted at the presence of an iron line
(Makishima et al. 1994).  However, it took a deep {\it ASCA} observation
to unambiguously detect the line and allow a detailed study (Reynolds,
Nowak \& Maloney 2000).  It was found that the line in NGC~4258 is
fairly weak (with an equivalent width of about 100\,eV) and narrow (with
a FWHM of less than $22000\kmps$).  Reynolds et al. (2000) argue that
this line does indeed originate from the accretion disk, implying that
the X-ray emitting corona has a size greater than $100GM/c^2$.  The
contrast between the iron lines found in NGC~4258 and its higher
luminosity Seyfert cousins is consistent with an ADAF scenerio for
LLAGNs.  However, the observational results are not yet conclusive.  If
the iron line seen in NGC~4258 comes from material not associated with
the accretion disk (such as a distant torus that is misaligned with the
almost edge-on accretion disk so as not to obscure the central engine
from our view), then the data are consistent with the presence of a
``Seyfert-like'' broad iron line.  See Reynolds et al. (2000) for
further details.

While it is significantly more luminous than NGC~4258, the well studied
active nucleus in the galaxy NGC~4051 is also often classified as a
LLAGN.  This object display a classic relativistic iron line indicating
the presence of a radiatively-efficient accretion disk in this object
(Guainazzi et al. 1996).  Wang et al. (1999) have recently discovered
interesting temporal variability in this iron line which displays
opposite trends to the variability found in MCG--60-30-15 --- both the
equivalent width and energy width of the line positively correlate with
the source flux.

\subsection{Radio-loud AGN}

It has long been known that most AGN can be readily characterized as
being radio-loud (i.e. with relativistic radio jets) or radio-quiet
(i.e. with no well-defined radio jets).  One of the greatest mysteries
in the field of AGN research is the physical mechanism underlying this
division.  A first step in solving this puzzle is to compare and
contrast the central engine structures of radio-quiet and radio-loud
AGN.  Since they originate from deep within the
central engine, X-rays are a good tool for probing any such differences.

Radio-loud AGN are rarer, and hence typically fainter, that their
radio-quiet counterparts.  Furthermore, many of the best candidates
for study are found in clusters of galaxies and it can be difficult to
observationally distinguish AGN emission from thermal cluster
emission. For these reasons, the quality of te observational
constraints is rather poorer in the case of radio-loud AGN as compared
with radio-quiet sources.  Having stated those caveats, there {\it
does} appear to be a difference between the X-ray properties of
radio-loud nuclei and radio-quiet nuclei.  Broad iron lines, and the
associated Compton reflection continua, are generally weak or absent
in the radio-loud counterparts (Eracleous, Halpern \& Livio 1996;
Wo{\'z}niak et al. 1997; Reynolds et al. 1997; Sambruna, Eracleous \&
Mushotzky 1999; Grandi et al 1999; Eracleous et al 2000). This effect
might be due to the swamping of a normal `Seyfert-like' X-ray spectrum
by a beamed jet component (similar to the swamping of optical emission
lines in a blazar spectrum).  Alternatively, the inner disk might be
in a physical state incapable of producing reflection signatures (such
as an ADAF or some similarly hot state).  Future observations with
{\it XMM--Newton} should be able to distinguish these possibilities by
searching for very weak broad components to the iron line.

\subsection{Galactic Black hole Candidates}

Smeared edges with little evidence of line emission have been observed
in the spectra of Galactic Black Hole Candidates (GBHC) (Ebisawa et al
1996). These observations can be explained if the surface of the disk is
moderately ionized with a mean ionization parameter of a few hundred
(Fig.~2 with $\xi=100-300$; Ross, Fabian \& Brandt 1996). As discussed in
Section 2.1, for this relatively narrow range of ionization parameters
line photons are resonantly trapped and eventually lost as Auger
electrons.  Hence, the X-ray reflection produces very little line
emission but an appreciable absorption edge. Sharp features will be
smeared as a result of the Doppler and relativistic effects, and this
blurring has possibly been detected in the spectrum of Cygnus~X--1 (Done
\& Zycki 1999).  Similar spectra but with broader emission and absorption
features are produced for higher values of the ionization parameter
$\xi\approxgt3\times10^3$, which appear to match those seen in GBHC.
The smearing in this instance is due to the line photons being generated
a few Thomson scattering depths into the disk (the very outermost layers
are completely ionized) and being Compton scattered on leaving it.

An interesting correlation has been claimed by Zdziarski, Lubi\`nski \&
Smith (1999) between the reflection fraction seen in accreting black
holes (AGN or GBHC) and the spectra index. Sources with flat spectra
tend to have a low reflection fraction. Models involving large central
holes in the disc, or ionized discs, may explain the correlation, as
may mild relativistic motion, thus beaming, of the continuum radiation
(Beloborodov 1999; see also Reynolds \& Fabian 1997).

It is currently difficult to discriminate between models in which a
cold disk truncates at a few tens of Schwarzschild radii (e.g.
Gierli\'nski et al 1997) and models in which an ionized disk extend in
to the innermost stable orbit (Young et al 2000) since both provide
good fits to present data. Future observations will hopefully resolve
these issues.

\section{Summary}

In this section we summarize some of the key points of this review.

\begin{list}{$\bullet$}{}
  
\item \emph{X-ray continuum.} The hard X-ray continuum in AGN and GBHCs
is thought to be produced in active or flaring regions in a corona above
the accretion disk. Thermal electrons multiply inverse Compton scatter
optical and UV photons from the disk to X-ray energies. The hard X-ray
power law that results irradiates the accretion disk and produces a
``reflection'' component in the spectrum.
  
\item \emph{Reflection component.} The reflection component causes the
  observed spectrum to flatten above 10\,keV as Compton recoil reduces
  the backscattered flux and also results in a strong iron fluorescence
  line at approximately 6.4\,keV. The precise energy and strength of the
  line depend on a number of factors such as the iron abundance, the
  inclination of the disk and its ionization state. An ionized disk may
  also produce a strong iron absorption edge.
  
\item \emph{Iron line profile.} The line profile is determined by Doppler
  shifts and relativistic boosting due to the motion of the disk and the
  gravitational redshifting of the black hole. This produces a broad,
  skewed line profile. Since the line originates from the innermost
  regions of the accretion disk, these effects are very pronounced.
  From observations of the line profile the black hole spin and the
  inclination angle of the accretion disk may be determined. In most
  Seyfert~1 galaxies (i.e. AGN in which we can view the accreting black
  hole directly) the accretion disk is inclined at about 30$^\circ$ to
  the observer. This is consistent with the standard model in which the
  Seyfert nucleus is surrounded by an optically thick torus with an
  opening angle of 30--40$^\circ$.

\item \emph{Observations.} The iron line was first clearly detected by
  \emph{Ginga} and a line profile subsequently resolved by \emph{ASCA}
  confirming the broad and skewed shape expected from an accretion disk
  around a Schwarzschild black hole. \emph{RXTE} has been able to study
  the line and continuum variability on much shorter timescales,
  although with reduced energy resolution.
  
\item \emph{Black hole spin.} The radius of the smallest stable circular
  orbit around the black hole decreases with the spin of the black
  hole. Since the line profile is sensitive to the innermost radius of
  fluorescent emission this may be used (with some assumptions about the
  astrophysics of this region) to estimate the spin of the black
  hole. With present time-averaged observations, however, such
  measurements may be ambiguous as alternative models with very
  different values for the black hole spin may produce almost identical
  line profiles.
  
\item \emph{Alternative models for the production of the broad iron
    line.} Models for the broad iron line that do not require a black
    hole accretion disk appear to fail. In particular the line width
    cannot be entirely due to Comptonization.  Hybrid models in which
    both Comptonization and Doppler/gravitational effects produce the
    line profile are heavily constrained.
  
\item \emph{Variability.} Rapid X-ray continuum variability is
  observed in most AGN and the iron line is expected to vary in response
  to this with a short time lag. Whilst these timescales are too short
  to be probed with present instruments, significant and complex iron
  line variability has been observed. Curiously, the line flux is seen
  to remain constant whilst the continuum changes, and there appears to
  be an anti-correlation between the reflected fraction and the
  equivalent width of the line.  In another study the reflected fraction
  and the photon index of the power law are correlated, both for an
  individual object, and between different objects (including both AGN
  and GBHC). Such observations need to be explained, especially since
  they appear contrary to our simple model of reflection.
  Flux-correlated changes in the ionization state of the disk may
  explain some of these facts.
  
\item \emph{Reverberation mapping.} The rapid X-ray variability is
  associated with the activation of new flares in the corona above the
  accretion disk. X-ray reverberation mapping is the technique of
  using observations of the iron line response, or ``echo'', to sudden
  changes in the continuum to study the accretion disk and black hole.
  In principle this may be used to determine the geometry of the X-ray
  emission and the black hole spin and mass. Such observations will be
  within the capabilities of the next generation of X-ray
  observatories.
  
\item \emph{Other classes of object.} Iron lines are observed in other
  classes of object in addition to Seyfert galaxies. In quasars the
  strength of the iron line decreases with increasing luminosity. This
  may be because the more luminous sources accreting closer to the
  Eddington limit and more highly ionized. The observation of iron lines
  in LLAGN may determine whether the accretion with low rates is an ADAF
  or a thin disk.  Weak iron lines have also been seen in LLAGN
  suggesting their low accretion rate flows are thin disks as opposed to
  geometrically thick ADAFs.  In radio loud AGN, broad iron lines and
  reflection humps are weak or absent, perhaps because the reflection
  signature is swamped by a beamed continuum.  All of these require
  further detailed observations. In GBHC the accretion disk is ionized
  and the reflection spectra show smeared absorption and emission
  feature, and there is debate as to the precise nature of the accretion
  flow within a few tens of Schwarzschild radii of the black hole.

\end{list}

Over the past decade observations of the broad iron line have provided
an unprecedented probe of the region within a few tens of Schwarzschild
radii of the black hole event horizon. The next generation of X-ray
observatories, beginning with \emph{XMM--Newton}, will address many of
the puzzling questions we have, and significantly enhance our
understanding of these enigmatic objects.

\section*{Acknowledgements}

We thank Mateusz Ruszkowski for comments on the manuscript.  ACF thanks
the Royal Society for support.  CSR appreciates support from Hubble
Fellowship grant HF-01113.01-98A.  This grant was awarded by the Space
Telescope Institute, which is operated by the Association of
Universities for Research in Astronomy, Inc., for NASA under contract
NAS 5-26555.

\end{document}